\documentclass[useAMS,usenatbib]{mn2e}
\usepackage{rotating}
\usepackage{journals}

\def\approxgt{\ifmmode \rlap{$>$}{}_{{}_{{}_{\textstyle\sim}}} \else%
$\rlap{$>$}{}_{{}_{{}_{\textstyle\sim}}}$\fi} 
\def\approxlt{\ifmmode \rlap{$<$}{}_{{}_{{}_{\textstyle\sim}}} \else%
$\rlap{$<$}{}_{{}_{{}_{\textstyle\sim}}}$\fi}

\def\farcs{\hbox{$.\!\!^{\prime\prime}$}}
\def\degr{\hbox{$^\circ$}}
\def\arcmin{\hbox{$^\prime$}}
\def\arcsec{\hbox{$^{\prime\prime}$}}

\LARGE \normalsize \title[1H~1905+000 in quiescence]{The neutron star soft X--ray
transient 1H~1905+000 in quiescence}

\author[Jonker et al.]  {P.G.~Jonker$^{1,2,3}$\thanks{email :
p.jonker@sron.nl. Based on observations made with ESO telescopes at the Paranal Observatories under
programme ID 073.D-0486(A)}, C.G. Bassa$^3$, G. Nelemans$^{4}$, A.M. Juett$^{5}$, E.F. Brown$^{6}$, D.
Chakrabarty$^7$ \\
$^1$SRON, National Institute for Space Research, Sorbonnelaan 2, 3584~CA, Utrecht, The Netherlands\\
$^2$Harvard--Smithsonian  Center for Astrophysics, 60 Garden Street, Cambridge, MA~02138, Massachusetts,
U.S.A.\\
$^3$Astronomical Institute, Utrecht University, P.O.Box 80000, 3508 TA, Utrecht, The Netherlands\\
$^4$Department of Astrophysics, IMAPP, Radboud University Nijmegen, Toernooiveld 1, 6525 ED, Nijmegen, The Netherlands\\
$^5$Department of Astronomy, University of Virginia, Charlottesville, VA~22903, U.S.A. \\
$^6$Department of Physics and Astronomy, Michigan State University, East Lansing, MI 48824, U.S.A.\\
$^7$Department of Physics and Kavli Institute for Astrophysics and Space Research, Massachusetts Institute
 of Technology, \\ Cambridge, MA 02139, U.S.A. \\
}

\begin{document}

\maketitle

\begin{abstract} \noindent In this Paper we report on our analysis of a $\sim25$ ksec.~{\it Chandra} X--ray
observation of the neutron star soft X--ray transient (SXT) 1H~1905+000 in quiescence. Furthermore, we discuss our
findings of the analysis of optical photometric observations which we obtained using the Magellan telescope and
photometric and spectroscopic observations which we obtained using the Very Large Telescope at Paranal. The X--ray
counterpart of 1H~1905+000 was not detected in our {\it Chandra} data, with a 95 per cent confidence limit to the
source count rate of 1.2$\times 10^{-4}$ counts s$^{-1}$. For different spectral models this yields an upper limit
on the luminosity  of 1.8$\times10^{31}$ erg s$^{-1}$ (for an upper limit on the distance of 10 kpc.) This
luminosity limit makes 1H~1905+000 the faintest neutron star SXT in quiescence observed to date. The neutron star
luminosity is so low that it is similar to the lowest luminosities derived for black hole SXTs in quiescence. This
low luminosity for a neutron star SXT challanges the hypothesis presented in the literature that black hole SXTs in
quiescence have lower luminosities than neutron star SXTs as a result of the presence of a black hole event
horizon. Furthermore, the limit on the neutron star luminosity obtained less than 20 years after the outburst has
ceased, constrains the thermal conductivity of the neutron star crust. Finally, the neutron star core must be so
cold that unless the time averaged mass accretion rate is lower than $2\times10^{-12}$ M$_\odot$ yr$^{-1}$, core
cooling has to proceed via enhanced neutrino emission processes. The time averaged mass accretion rate can be
derived from binary evolution models if the orbital period of the system is known. Our optical observations show
that the optical counterpart discovered when the source was in outburst has faded. Near the outburst optical
position we find two stars with a separation of 0.7\arcsec and I=19.3$\pm0.1$ and 21.3$\pm0.1$. VLT optical
spectroscopy revealed that the spectrum of the brighter of the two sources is a G5--7V--star. However, the outburst
astrometric position of the optical counterpart does not coincide with the position of the G5--7V--star nor with
that of the fainter star. We derive a limit on the absolute I--band magnitude of the quiescent counterpart of
M$_I>$7.8 assuming the source is at 10 kpc. This is in line with  1H~1905+000 being an ultra--compact X--ray
binary, as has been proposed based on the low outburst V--band absolute magnitude.

\end{abstract}


\section{Introduction} 

Low--mass X--ray binaries are binary systems in which a $\approxlt1 M_{\odot}$ star transfers matter to a neutron
star or a black hole. A large fraction of the low--mass X--ray binaries are transient systems -- the so called soft
X--ray transients (SXTs; e.g.~see \citealt{1997ApJ...491..312C}). Before the launch of the XMM--Newton and {\it
Chandra} satellites only a few (mostly) nearby SXTs could be studied in quiescence (e.g.~the black hole candidates
A~0620--00 and V404~Cyg and the neutron star systems Cen~X--4 and Aql~X--1; \citealt{1994ApJ...429L..25W};
\citealt{1995ApJ...442..358M}; \citealt{1987A&A...182...47V}). Using the XMM--Newton and {\it Chandra} satellites
many more systems were studied in quiescence in the initial years of operation (see
e.g.~\citealt{2001ApJ...553L..47G}, \citealt{2002ApJ...570..277K}, \citealt{2002ApJ...580..413R},
\citealt{2001ApJ...560L.159W}, \citealt{2002ApJ...575L..15C}, \citealt{2003A&A...399..631H},
\citealt{2004MNRAS.354..666J}). Contemporaneous theoretical progress provided the framework for the interpretation
of these observations (\citealt{1994ApJ...428L..13N}, \citealt{1997ApJ...478L..79N}; \citealt{1998ApJ...504L..95B},
\citealt{2001ApJ...548L.175C}; \citealt{1996A&A...315..141Z}; \citealt{2002A&A...386.1001G}) which turned out to
have a profound impact on two important areas of high energy astrophysics.

First, comparing the quiescent luminosity of neutron star SXTs with that of black hole SXTs it was found
that black hole (BH) SXTs are systematically fainter in quiescence than neutron stars
(\citealt{1997ApJ...478L..79N}, \citealt{1999ApJ...520..276M}, \citealt{2001ApJ...553L..47G},
\citealt{2002ApJ...570..277K}). This has been interpreted as evidence for advection of energy across a BH
event horizon. If true this would constitute the first confirmation of a prediction of Einstein's Theory of
General Relativity in the strong field regime. Despite many objections to this interpretation
(\citealt{2000ApJ...541..849C}; \citealt{2002A&A...396L..31A}), alternative explanations for the difference
in quiescent luminosity (\citealt{2003MNRAS.343L..99F}), and neutron stars which turned out to be  fainter
than initially found to be the rule (e.g.~SAX~J1808.4--3658, \citealt{2002ApJ...575L..15C}; EXO~1747--214;
Tomsick et al.~2005), none of the neutron star SXTs have quiescent luminosities as low as the faintest BH
SXTs, which have 0.5--10 keV luminosities   $<10^{31}$ erg s$^{-1}$; (e.g.~\citealt{2002ApJ...570..277K},
\citealt{2003A&A...399..631H}). Hence, irrespective of the interpretation, the difference in quiescent
luminosity between BH and neutron star SXTs seems to be one of the very few distinct characteristics
between BHs and neutron stars.

Secondly, the quiescent spectra of neutron star SXTs are well--fit by a neutron star atmosphere model (NSA)
sometimes supplemented with a power--law component. Especially in sources with a quiescent luminosity near
$10^{33}$ erg s$^{-1}$ the spectrum is dominated by a strong thermal component (\citealt{2004MNRAS.354..666J}).
The thermal component is thought to be due to the hot neutron star core moderated by the neutron star
atmosphere. The neutron star core temperature can be calculated by combining well established theories about
the time--averaged mass accretion rates in neutron star SXTs (\citealt{1962ApJ...136..312K};
\citealt{verbunt1995}), the pycnonuclear reactions taking place in the neutron star crust
(\citealt{1969ApJ...155..183S}; \citealt{1990A&A...227..431H}; \citealt{2000ApJ...539..888K}) and theoretical
neutron star cooling predictions (see \citealt{2004ARA&A..42..169Y} for a review).  Therefore, in theory, an
NSA--fit provides means to measure the mass and radius of the neutron star and hence constrain the equation of
state (EoS) of matter at supranuclear densities. The description of the relations between pressure and density
of matter (the EoS) under the extreme conditions encountered in neutron stars is one of the ultimate goals of
the study of neutron stars.

In practice, numbers typical for a canonical neutron star were found (e.g.~\citealt{2003ApJ...598..501H}),
rendering support for this interpretation. However, there is an ongoing debate whether the temperature of
the thermal (NSA) component is varying on short timescales  (cf.~\citealt{2002ApJ...577..346R},
\citealt{2004ApJ...601..474C}, and \citealt{igrj00291jonker}). Small temperature changes could be
explained by changes in the neutron star atmosphere due to ongoing low--level accretion
(\citealt{2002ApJ...574..920B}). Large changes on short timescales would render it unlikely that the
soft/thermal component is due to cooling of the neutron star, limiting the applicability of the NSA model
fit. Finally, there are currently two sources known which returned to quiescence after a several
year--long accretion epoch (i.e.~KS~1731--260 and MXB~1659--298). As a result of these long accretion
episodes the neutron star {\it crust} is heated to temperatures larger than that of the core. The observed
thermal spectral component has been identified as cooling of the neutron star crust
(\citealt{2002ApJ...573L..45W}; \citealt{2002ApJ...580..413R}). As the crust cools the X--ray spectral
properties also change slightly (\citealt{2004wijninpress}).

Recent {\it Chandra} observations of accretion powered millisecond X--ray pulsars in quiescence found that
the quiescent luminosity of many of those observed so far, not just SAX~J1808.4--3658, is low
(\citealt{2005ApJ...619..492W}; \citealt{2005A&A...434L...9C}). A possible exception could be the
accretion--powered millisecond X--ray pulsar IGR~J00291+5934 (\citealt{igrj00291jonker}). Furthermore, the
X--ray spectrum is in most cases dominated by a power--law component similar to that of quiescent BH
(\citealt{2005ApJ...619..492W}). Hence, the dichotomy between the BH and neutron star quiescent luminosity
may not be as large as previously derived (see also \citealt{2004MNRAS.354..355J}). We note however, that
reliable distance estimates could be made for only 2 accretion powered millisecond systems,
SAX~J1808.4--3658 and XTE~J1814--338 (\citealt{2001A&A...372..916I}; \citealt{2003ApJ...596L..67S}). For
the other systems the distance estimates are rather uncertain, making the quiescent luminosity uncertain as
well. The low--luminosity and the small contribution of a thermal spectral component to the luminosity of
SAX~J1808.4--3658 ($<$10\%; \citealt{2002ApJ...575L..15C}, although see the comment about this upper limit
in \citealt{2004pycnoyakov}) hint at a massive neutron star ($M>1.7 M_\odot$;
\citealt{2003A&A...407..265Y}, \citealt{2004ARA&A..42..169Y}). The upper limit on the thermal spectral
component implies that the neutron star core of SAX~J1808.4--3658 must release the energy produced in the
crust due to pycnonuclear reactions rapidly via enhanced neutrino emission. This enhanced neutrino emission
can only occur when the neutron star mass is larger than the canonical 1.4 $\,M_\odot$. 

1H~1905+000 was first detected on MJD~42368 (UTC) by Ariel~5 (\citealt{1976MNRAS.175P..39S}). Six type I X--ray
bursts were discovered on different occasions by SAS--3 firmly establishing the nature of the compact object as a
neutron star (\citealt{1976MNRAS.177P..93L}). The last reported detection of the source was that by EXOSAT on
MJD~46316 (UTC). A radius expansion burst was detected on this occasion (\citealt{1990A&A...228..115C}). During the
period of activity the source has also been detected with HEAO--1 and Einstein (\citealt{1980AJ.....85.1062R} and
\citealt{1997ApJS..109..177C}, respectively). However, the source was not detected in the ROSAT All Sky Survey
(\citealt{2005ApJ...627..926J}). The source likely went to quiescence at the end of the 1980s/early 1990s. Next,
1H~1905+000 was observed for 5 ks with the back--illuminated S3 CCD--chip of the Advanced CCD Imaging Spectrometer
(ACIS) detector on board the {\it Chandra} satellite with the High--Energy Transmission Grating inserted
(\citealt{2005ApJ...627..926J}). Again no source was detected at the position of the optical counterpart discovered
when the source was in outburst (\citealt{1985A&A...147L...3C}). The derived upper limit on the unabsorbed 0.5--10
keV  flux for 1H~1905+000 was  1$\times10^{-14}$ erg cm$^{-2}$ s$^{-1}$ for an assumed black body spectrum with a
temperature of 0.3 keV. The distance for 1H~1905+000 derived from the observed radius expansion burst peak flux is
7.3 or 10 kpc (\citealt{2004MNRAS.354..355J}; the values assume hydrogen and helium bursts, respectively). From
Einstein observations \citet{1997ApJS..109..177C} determined that the interstellar extinction, N$_H$, to
1H~1905+000 is $(1.9\pm0.2)\times10^{21}$ cm$^{-2}$. This yields an upper limit to the intrinsic (i.e.~corrected
for the interstellar extinction) 0.5--10 keV source luminosity of $1.0-1.7\times10^{32}$ erg s$^{-1}$ for
1H~1905+000. In summary, it is likely that the source had been accreting steadily at L$\sim4\times10^{36}$ erg
s$^{-1}$ for more than 10 years before returning to quiescence. In this Paper we present our analysis of a $\sim25$
ksec.~{\it Chandra} observation of this neutron star SXT in quiescence. Furthermore, our analysis of Very Large
Telescope and Magellan optical observations of the region of the source in quiescence is also presented.

\section{Observations, analysis and results} 

\subsection{Optical Magellan, VLT, archival WHT and CFHT observations}

\begin{figure*}
  \includegraphics[angle=0,width=\textwidth,clip]{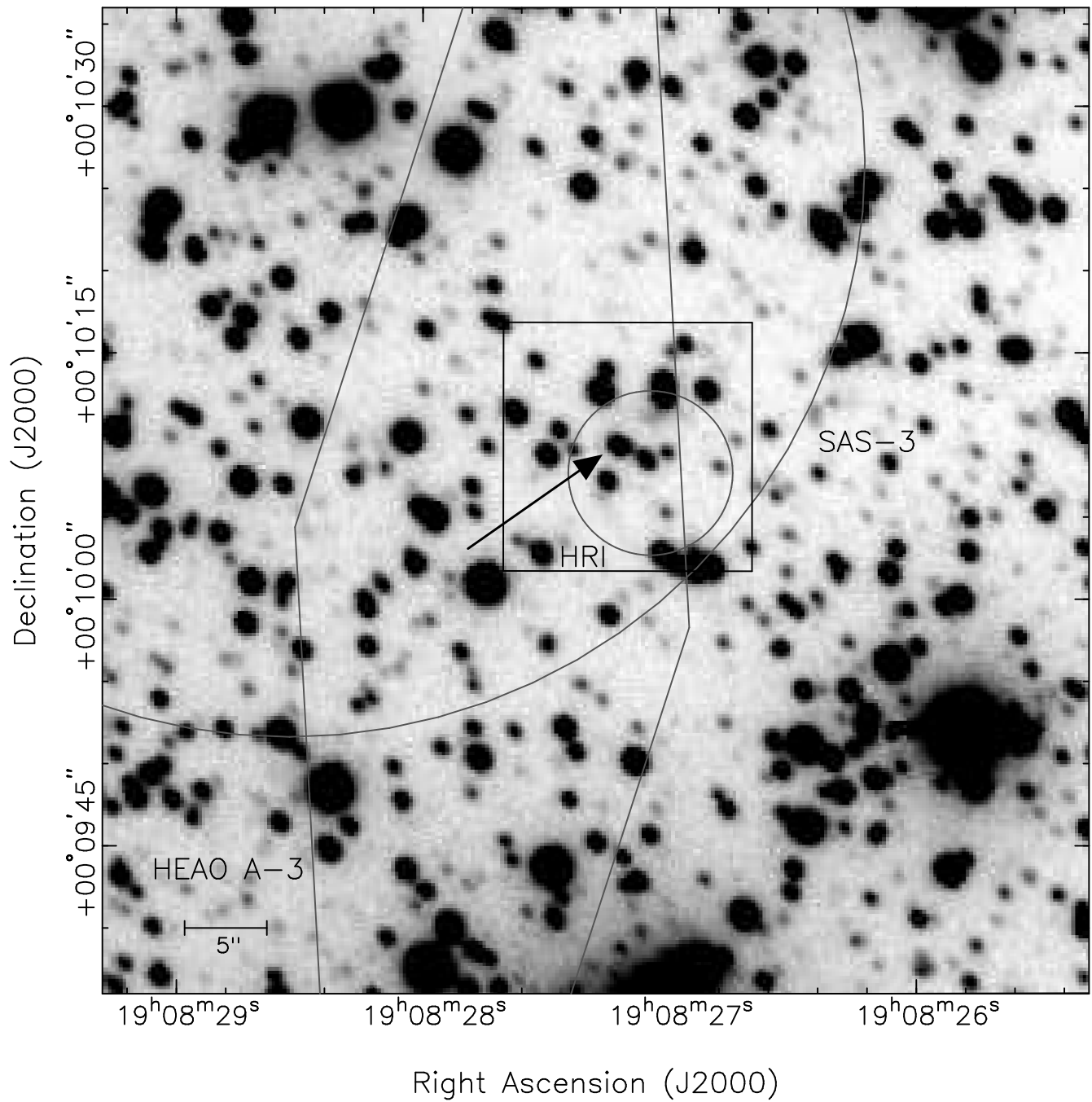}
  
  \caption{An I--band finder chart of the field of 1H~1905+000 obtained by median combining the two 300 seconds
  images obtained with Magellan/IMACS (1'x1', North is up and East is left) . Overplotted are the HEAO A--3
  (diamond shape), the Einstein HRI (small circle) and the SAS--3 (large circle) error regions. The box shows the
  region  plotted in Figure 2. The arrow indicates the approximate position of the blue counterpart discovered when
  1H~1905+000 was in outburst (\citealt{1985A&A...147L...3C}.) Psf--fitting showed that this star consists of two
  stars separated by 0.7\arcsec (see Figure 2). }

\label{fig:src}
\end{figure*}

\begin{figure*}
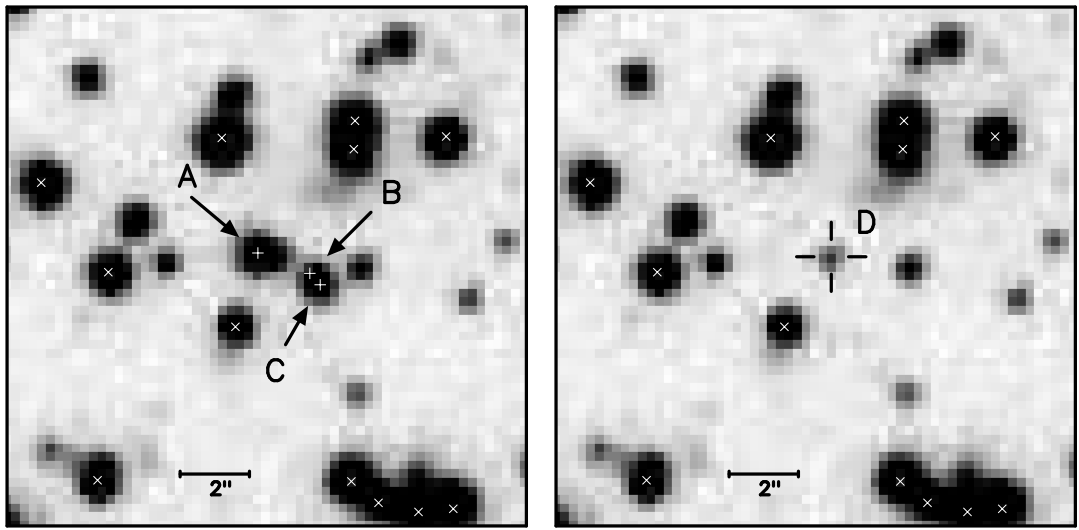

  \includegraphics[angle=0,width=7cm]{on.ps}\quad
  \includegraphics[angle=0,width=7cm]{off.ps}
  
  \caption{{\it Left:} A median combined image of six 10 seconds VLT FORS2 white light observations of
  1H~1905+000 obtained under excellent seeing conditions ($<0.6$\arcsec). {\it Right:} The three stars (A,B,C)
  indicated with a small, white plus sign in the {\it left} image have been subtracted using
  psf--fitting. A star very close to the position of the optical counterpart in outburst remains (we
  call this star D).}

\label{fig:onoff}
\end{figure*}

In order to determine the best (optical) position of 1H~1905+000 and to search for the optical counterpart in
quiescence, we have obtained I--band images with exposure times of 10~seconds and 2x300 seconds using the
Inamori--Magellan Areal Camera and Spectrograph (IMACS) instrument mounted on the 6.5~m Magellan--Baade telescope
on July 7, 2005, 03:03 UTC (MJD 53558.14594 UTC). The seeing was 0.69\arcsec. Using the second USNO CCD
Astrograph Catalog (UCAC2) catalogue (\citealt{2004AJ....127.3043Z}) we determined the position of 71 bright,
unsaturated, stars in the 10--second IMACS image to obtain an astrometric solution (the rms of the fit was
0\farcs060 both in right ascension [$\alpha$] and in declination [$\delta$]). Subsequently, the astrometric
solution of the 10--second frame was transferred to the 300--second images using 1397 stars. In this the
uncertainty was 0\farcs017 in $\alpha$ and 0\farcs015 in $\delta$. Hence, the absolute uncertainty in the optical
astrometry of the 300 second images is 0\farcs062 in $\alpha$ and 0\farcs061 in $\delta$. Next, standard image
processing was done in \textsc {midas} (i.e.~bias subtraction and flatfield correction.) The two 300~seconds
observations were averaged (see Fig.~\ref{fig:src}). We have observed standard stars on the same CCD close in
time and airmass to the IMACS observations of 1H~1905+000. Using point spread function fitting (psf--fitting)
techniques we found that the star present near the optical position of the counterpart discovered in outburst
consists of two stars close together (within 0.7\arcsec) with I--band magnitudes 19.3$\pm0.1$ (star A) and
21.3$\pm0.1$ (star D; see Fig.~\ref{fig:onoff}). Star A has a position $\alpha_{J2000.0}=19{\rm ^h}$08${\rm
^m}$27${\rm ^s}$.217$\pm$0\farcs063, $\delta_{J2000.0}= +00^\circ10'$09\farcs42$\pm$0\farcs062. Star D has a
position $\alpha_{J2000.0}= {\rm 19^h 08^m 27^s.171}\pm$0\farcs063, $\delta_{J2000.0}=
+00^\circ10'09$\farcs29$\pm$0\farcs062 (68 per cent confidence uncertainty; the uncertainty in this position is
the square root of the quadratically added internal uncertainty [0\farcs01 in both $\alpha$ and $\delta$] and the
uncertainty in the absolute calibration of the astrometric solution mentioned earlier). As a (conservative) limit
on the detection limit of the 2x300 second I--band image we determined the magnitude of the faintest star
detected at 5 $\sigma$; it has I=23.5. 

We have also obtained 14 white light images with the FOcal Reducer and low dispersion Spectrograph 2 (FORS2)
mounted on the 8.2~m Very Large Telescope (VLT) Yepun (images were obtained on MJD 53135.3668, 53135.3714,
53143.3471, 53143.3504, 53144.1639, 53146.3858, 53146.3901, 53148.2501, 53148.2555, 53148.2656, 53148.2665,
53148.302, 53148.3606, 53148.3624 UTC). Each of these images has an exposure time of 10 seconds. We corrected for
bias using the overscan region of the CCD, however, no white light flatfield images are available since these
images were acquisition images for spectroscopic observations (see below). Therefore, we could not correct for
pixel--to--pixel variations in sensitivity. We used psf--fitting in order to determine the relative brightness of
the two stars present close to the position of the optical counterpart in outburst (see Fig.~\ref{fig:onoff}). We
were able to use 12 out of the 14 acquisition observations for which the seeing conditions were
0.45\arcsec--0.82\arcsec to search for white light variability. The rms scatter in the magnitude of star D is 0.11
magnitudes. However, this variability could have been introduced by the psf--fitting technique since the rms
variability in the fainter of the two stars in another star--pair of similar brightness ratio and separation was
0.09 magnitudes. We conclude that star D did not vary significantly over the course of our observations. We median
combined six of the 10 seconds images with the best seeing (seeing $<$0.6\arcsec). Next, we again used the UCAC2
catalogue (\citealt{2004AJ....127.3043Z}) to determine the position of 23 bright, unsaturated, stars in the
resultant image to obtain an astrometric solution (the rms of the fit was 0\farcs063 in $\alpha$ and 0\farcs082 in
$\delta$. The position of star A and D are consistent with being the same as during the IMACS observations.

\begin{figure*}
  \includegraphics[angle=0,width=14cm]{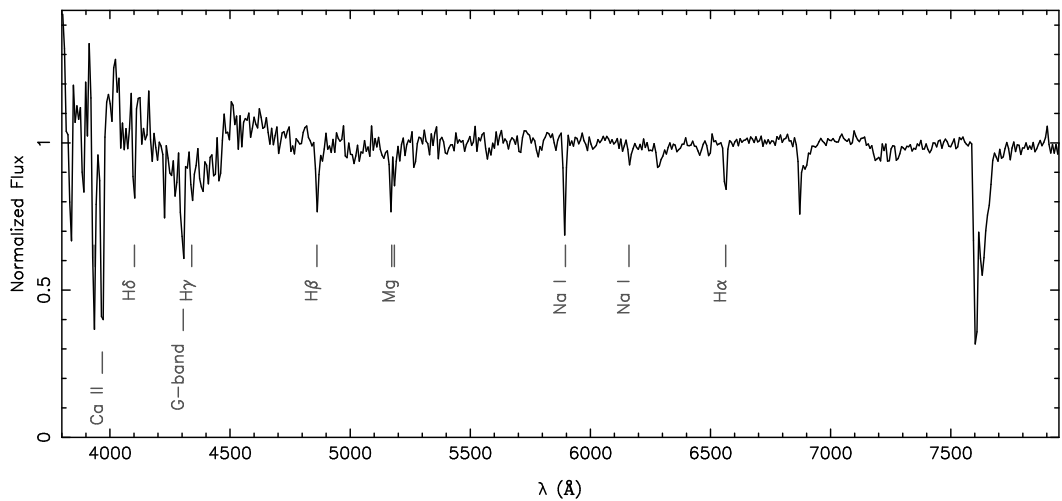}
  
  \caption{The combined, normalised VLT FORS2 600B (2.3 hours total exposure) and 600RI (3.05 hours
  total exposure) grating spectrum of the star at the position of the outburst optical counterpart of
  1H~1905+000. The spectrum resembles that of a G5--7V--star.}

\label{fig:spec}
\end{figure*}

As mentioned above the white light images are acquisition images for spectroscopic observations. We have obtained
VLT/FORS2 spectra of star A (cf.~Fig.~\ref{fig:src} \&~\ref{fig:onoff}) using the 600B and 600RI gratings with an
exposure time of $\sim$2750 seconds on MJD 53146.4006, 53148.3059, 53148.3659 and MJD 53135.3744, 53143.3541,
53143.3883, 53148.2690, respectively. Hence, the total exposure in the 600B grating spectrum was $\sim$2.3 hours
and in the 600RI grating spectrum it was $\sim$3.05 hours. A slit width of 1\arcsec was used on each occasion. The
dispersion was 1.5\AA~pixel$^{-1}$ at 4429\AA~with the 600B grating and 1.65\AA~pixel$^{-1}$ at 6552\AA~with the
600RI grating. With a slit width of 1\arcsec, the spectral resolution varies from approximately 400 km s$^{-1}$ at
4430\AA~to 300 km s$^{-1}$ at 6550\AA. The spectra were extracted and reduced using \textsc {iraf}\footnote{\textsc
{iraf} is distributed by the National Optical Astronomy Observatories}. Once the spectra were reduced the spectral
analysis was done using \textsc {molly}. The spectrum of this star is consistent with a G5--7V star (see
Fig.~\ref{fig:spec}). 

We have median combined 14 archival V--band observations of the field of 1H~1905+000 obtained with the AUX port
camera mounted on the 4.2~m William Herschel Telescope located at the Roque de Los Muchachos Observatory, La
Palma, Spain on July 30, 1994 (MJD 49563 UTC). These images have been retrieved from the ING Archive. In total
100 images had been obtained but we only selected the 14 with best seeing conditions (seeing $<$0.8\arcsec).
The G--star is detected at a magnitude V=20.61$\pm0.01$ (statistical error only) and star D is barely detected
at V=23.3$\pm0.1$. We note however that only one standard star was observed and only one filter was obtained.
Hence, colour corrections could be important (these systematic uncertainties are not included). The positions
of both sources is consistent with that derived from the IMACS images.

Finally, we have obtained a subsection ($37\arcsec\times37\arcsec$) of the 1984 $V$-band outburst
Canada--France--Hawaii Telescope (CFHT) image published by Chevalier et al.~(1985; Ilovaisky 2005, priv.\ comm.).
We have astrometrically tied this image to the 10\,s IMACS $I$-band image. The uncertainty in the tie is 0\farcs050
in $\alpha$ and 0\farcs055 in $\delta$. Although the resolution of the image is  worse than that of the FORS2 and
IMACS images, it is clear that there is excess emission compared to that contributable to stars at the position of
star A and D (Fig.~5). Using a Gaussian for the psf of the outburst image, we have fitted the stars on the image
with the aim to determine the position of the source in outburst taking into account the flux from star A and D.
The positions of the stars in the IMACS $I$--band observations were transformed to the CFHT image and kept fixed
during the fitting process. As such, we only fitted for the overall background, the fluxes of the IMACS stars and
the flux and position of the source in outburst. For the latter we obtain
$\alpha=19^\mathrm{h}08^\mathrm{m}27\fs200$, $\delta=+00\degr10\arcmin09\farcs10$. Here, the intrinsic uncertainty
on the source position is 0\farcs03 in both $\alpha$ and $\delta$. For the absolute uncertainty this should be
quadratically added to the uncertainties of the ties between the CFHT $V$--band and IMACS $I$--band image (see
above) and the IMACS $I$--band image and the UCAC2 catalog (0\farcs060 in both $\alpha$ and $\delta$). However, for
comparison of the outburst position with that of star A and D, we can directly compare the outburst position with
the positions on the 5\,min IMACS $I$--band image. For this, we can neglect the IMACS--UCAC2 uncertainty, but must
include the uncertainty in the tie between the 10\,sec and 5\,min IMACS images (0\farcs017 in $\alpha$ and
0\farcs015 in $\delta$). As such, star A is offset from the outburst position by $-0\farcs255\pm0\farcs062$ in
$\alpha$ and $-0\farcs320\pm0\farcs065$ in $\delta$, while star D is offset by $0\farcs435\pm0\farcs062$ and
$-0\farcs190\pm0\farcs065$. These offsets correspond to $4.5\sigma$ for star A and $5.3\sigma$ for star D. In
Fig.~\ref{fig:old} we have overplotted with small crosses the position of several reference stars detected in
Fig.~\ref{fig:onoff} and with a small circle the position of the outburst optical counterpart. As can be seen it is
unlikely that the outburst counterpart can be associated with star D unless that star has a high proper motion of
24 milliarcseconds year$^{-1}$. At a distance of 10 kpc this would convert into a rather large velocity of
$\sim1100$ km s$^{-1}$. Finally, we have investigated whether the differential Galactic rotation at
l$^{ii}=35^\circ$ can be used to explain the observed offset of star D from the outburst optical position, under
the assumption that star D arises due to the companion star and/or accretion disc of 1H~1905+000 at 10 kpc or at
7.5 kpc. The change in position with respect to other field stars is less than 0.1\arcsec over the 20 year that
separate the CFHT V--band outburst observations and the Magellan I--band observations. This is insufficient to
explain the observed offset. We conclude that star D is not the quiescent optical counterpart to 1H~1905+000. 

\begin{figure*}
  \includegraphics[angle=-90,width=7cm]{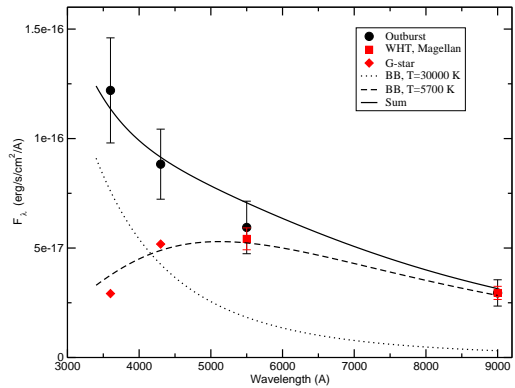}
  
  \caption{The optical spectral energy distribution of the outburst source and the G5--7 star. As can be seen
  the outburst SED can be explained as the superposition (solid line) of a hot, small black body component
  (T=30000 K, R=0.08 R$_\odot$; dotted line) and the G--star (T=5700 K, R=1 R$_\odot$; dashed line). Note that
  the lines (dotted, dashed and drawn) do not represent formal fits to the data points. The circles with error
  bars represent the dereddened optical outburst measurements (CFHT measurements; Chevalier et al.~1985). The
  squares are the dereddened observed magnitudes corresponding to the G--star when the low--mass X--ray binary
  was in quiescence (WHT and Magellan measurements; this work). The diamonds are the dereddened magnitudes
  derived using the observed spectral type and the source distance necessary to explain the G--star I--band
  magnitude.  }

\label{fig:sed}
\end{figure*}

As mentioned above, the G--star contributes significantly to the outburst V--band magnitude measured by
\citet{1985A&A...147L...3C}. We used the properties of the G5--7V star and the V and I--band magnitudes observed
when the low--mass X--ray binary was in quiescence to determine the G--star distance. In this we follow 
\citet{1985A&A...147L...3C} who noted that the interstellar extinction does not increase significantly in the
direction of 1H~1905+000 for sources with a distance larger than 4 kpc. Hence, we used the same N$_H$ for the
G--star as was found for 1H~1905+000 in outburst. We used \citet{1985ApJ...288..618R} to convert N$_H$ to an A$_V$
and the tables of \citet{scfida1998} to convert A$_V$ to A$_U$, A$_B$ and A$_I$. We corrected the observed
magnitudes for the interstellar extinction and plotted the optical SED for the outburst source as well as for the
G--star (see Figure \ref{fig:sed}). To indicate the contribution of the  outburst accretion disc, we included in
the plot the SED contribution of a small, spherical, hot component. For the G--star we find a distance of 8.5 kpc
(fixing the radius to 1~R$_\odot$), for the accretion disc we took 10 kpc for its distance and we get a radius of
$\approx$0.08 R$_\odot$ for a temperature of 3$\times10^4$ K. Such a temperature is in the range of temperatures
found for accretion discs around low--mass X--ray binaries (see for instance \citealt{1995xrb..book...58V} and
references therein). The fact that only such a small disc can be accommodated adds to the evidence that 1H~1905+000
is an ultra--compact X--ray binary.

\begin{figure*}
  \includegraphics[angle=0,width=7cm]{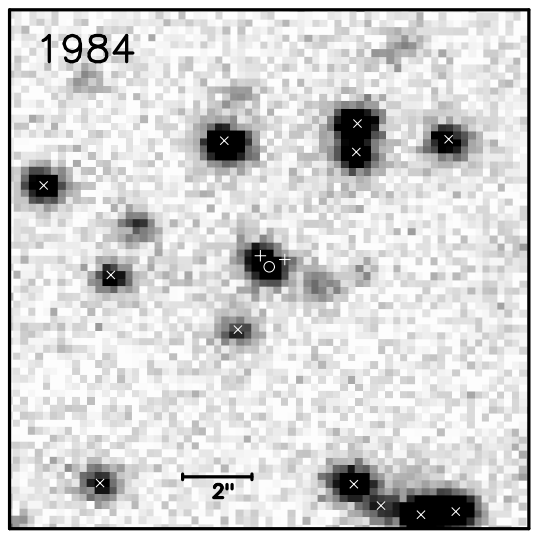}
  
  \caption{The 1984 CFHT V--band observation presented in Chevalier et al.~(1985) with crosses overplotted on the
  reference stars and "+" signs on the G--star and star D. The outburst optical position depicted by a small circle
  is offset from both the centroid position of the G--star and that of star D.}

\label{fig:old}
\end{figure*}

\subsection{{\it Chandra} X--ray observations}

We observed 1H~1905+000 with the back--illuminated S3 CCD--chip of the Advanced CCD Imaging Spectrometer
(ACIS) detector on board the {\it Chandra} satellite. The observations started on MJD~53425.852665 (UTC;
Feb.~24, 2005). The net, on--source exposure time was $\sim$24.8 ks. The data telemetry mode was set to
{\sl very faint} to allow for a better background subtraction. After the data were processed by the {\it
Chandra} X--ray Center (ASCDS version 7.5.0), we analysed them using the {\it CIAO 3.2.1} software
developed by the Chandra X--ray Center. We reprocessed the data to clean the background and take full
advantage of the {\sl very faint} data mode. We searched the data for background flares but none were
found, hence we used all data in our analysis. We detect three sources in the field of view of the ACIS S3
CCD. 

Since one of the detected X--ray sources (source 1 below) has an optical counterpart detectable in our
FORS2 white light images, we use the accurate optical position of this source to apply a boresight
correction to the {\it Chandra} observation and hence improve the astrometric accuracy of the {\it Chandra}
observation. The boresight shift that we find is: $\Delta\alpha$=-0\farcs210$\pm0\farcs071$,
$\Delta\delta=+0\farcs080\pm0\farcs088$. The J2000.0 $\alpha$ and $\delta$ of the three detected X--ray
sources are:\\(1) $\alpha_{J2000.0}=19{\rm ^h08^m34^s.094}$, $\delta_{J2000.0}= +00^\circ11'39$\farcs09
(with an error of 0\farcs078 in $\alpha$ and 0\farcs094 in $\delta$)\\(2) $\alpha_{J2000.0}{\rm
=19^h08^m22^s.183}$, $\delta_{J2000.0}=+00^\circ07'33$\farcs65 (with an error of 0\farcs11 in $\alpha$ and
0\farcs12 in $\delta$)\\(3) $\alpha_{J2000.0}={\rm 19^h08^19^s.878}$, $\delta_{J2000.0}=
+00^\circ06'18$\farcs20 (with an error of 0\farcs13 in $\alpha$ and 0\farcs15 in $\delta$).\\ We assign the
following names to these sources CXOU~J190834.1+001139, CXOU~J190822.2+000734, CXOU~J190819.9+000618,
respectively. The X--ray source closest to the position of the optical counterpart found in outburst
(\citealt{1985A&A...147L...3C}) is more than 2 arcminutes away. We do not detect a source at the position
of the optical outburst source as measured in the CFHT V--band image in our $\sim$25 ks--long {\it Chandra}
observation. Furthermore, we detect no X--ray photons within a 1\arcsec circle centred on the optical
outburst position. Following \citet{1986ApJ...303..336G}, we take an upper limit of 3 source photons to
determine the $\sim$95 per cent upper limit on the source count rate of $1.2\times 10^{-4}$ counts
s$^{-1}$. We used PIMMS version 3.6a\footnote{available at http://cxc.harvard.edu/toolkit/pimms.jsp} to
estimate upper limits on the source flux for a given interstellar extinction and an (assumed) spectral
energy distribution for the source. In Table~\ref{UL} we give these upper limits to the unabsorbed 0.5--10
keV source flux and 0.5--10 keV source luminosity.

\begin{table*}

\caption{Upper limits to the unabsorbed 0.5--10 keV source flux (F) and 0.5--10 keV luminosity (L) for
various values of the interstellar extinction, different spectral energy distributions of the source,
and using 7.5 and 10 kpc for the distance of 1H~1905+000. PL stands for power law, BB for blackbody.}

\label{UL}
\begin{tabular}{ccccc}
\hline
${\rm N_H}$ & Model & F${\rm _{0.5-10\,keV}}$ unabsorbed & L${\rm _{0.5-10\,keV, d=10 kpc}}$ & L${\rm_{0.5-10 keV, d=7.5 kpc}}$\\
cm$^{-2}$   & & erg cm$^{-2}$ s$^{-1}$ & erg s$^{-1}$ & erg s$^{-1}$ \\
\hline
\hline
& & 1H~1905+000 & & \\
\hline
2.1$\times10^{21}$   & PL index=2.0 &  1.2$\times 10^{-15}$& 1.4$\times 10^{31}$& 8.1$\times 10^{30}$  \\
2.1$\times10^{21}$   & PL index=1.5 &  1.5$\times 10^{-15}$& 1.8$\times 10^{31}$& 1.0$\times 10^{31}$\\
2.1$\times10^{21}$   & BB temperature=0.2 keV &  8.3$\times 10^{-16}$& 9.9$\times 10^{30}$& 5.6$\times 10^{30}$\\
2.1$\times10^{21}$   & BB temperature=0.3 keV&  7.3$\times 10^{-16}$& 8.7$\times 10^{30}$& 4.9$\times 10^{30}$\\

1.9$\times10^{21}$   & PL index=2.0 &  1.2$\times 10^{-15}$& 1.4$\times 10^{31}$ & 7.8$\times 10^{30}$\\
1.9$\times10^{21}$   & PL index=1.5 &  1.5$\times 10^{-15}$& 1.8$\times 10^{31}$ & 1.0$\times 10^{31}$\\
1.9$\times10^{21}$   & BB temperature=0.2 keV&  7.8$\times 10^{-16}$& 9.4$\times 10^{30}$ & 5.3$\times 10^{30}$\\
1.9$\times10^{21}$   & BB temperature=0.3 keV&  7.0$\times 10^{-16}$& 8.4$\times 10^{30}$ & 4.7$\times 10^{30}$\\

1.7$\times10^{21}$   & PL index=2.0 &  1.1$\times 10^{-15}$& 1.3$\times 10^{31}$ & 7.5$\times 10^{30}$\\
1.7$\times10^{21}$   & PL index=1.5 &  1.5$\times 10^{-15}$& 1.7$\times 10^{31}$   & 9.8$\times 10^{30}$\\
1.7$\times10^{21}$   & BB temperature=0.2 keV&  7.4$\times 10^{-16}$& 8.8$\times 10^{30}$ & 5.0$\times 10^{30}$\\
1.7$\times10^{21}$   & BB temperature=0.3 keV&  6.7$\times 10^{-16}$& 8.0$\times 10^{30}$ & 4.5$\times 10^{30}$\\

\end{tabular}
\end{table*}

\section{Discussion}

We have observed the field of the neutron star SXT 1H~1905+000 in X--rays with the {\it Chandra} satellite for
$\sim25$ ksec, but did not detect the source in quiescence. Depending on the assumed spectral energy distribution,
the interstellar extinction, and the distance we derive an upper limit to the 0.5--10 keV luminosity for
1H~1905+000 of ${\rm L_{0.5-10 keV, d=10 kpc}<1.8 \times 10^{31} erg s^{-1}}$ whereas, ${\rm L_{0.5-10 keV, d=7.5
kpc}<1.0 \times 10^{31} erg s^{-1}}$. If we assume a neutron star of radius 10 km and a neutron star atmosphere
model as given by \citet{1996A&A...315..141Z} then the upper limit on the 0.5--10 keV luminosity can be converted
to an upper limit on the effective temperature of ${\rm T_{eff}< 7.5\times10^5}$ K. These upper limits imply that
the quiescent X--ray luminosity of 1H~1905+000 is the lowest of any neutron star SXT observed so far for which
there is a reliable distance estimate. It was found earlier that the outburst absolute V--band magnitude of the
system was 4 (assuming a distance of 10 kpc; \citealt{1985A&A...147L...3C}). Compared with other SXTs in outburst
this is rather low, which is typical for systems with orbital periods less than $\sim$80 minutes (also known as
ultra--compact systems; see \citealt{vamc1994}). Furthermore, from our optical Magellan/IMACS I--band observations
of the system in quiescence and the precise astrometry we conclude that we do not detect the optical counterpart of
1H~1905+000 in quiescence, which for a distance of 10 kpc gives a limit on the absolute I--band magnitude of
1H~1905+000 in quiescence of M$_I>7.8$. In this we took N$_H=1.9\times10^{21}$ cm$^{-2}$ as found during outburst, used
\citet{1985ApJ...288..618R} to convert N$_H$ to an A$_V$ and the tables of \citet{scfida1998} to convert A$_V$ to
A$_I$. Our conservative upper limit on the quiescent optical absolute magnitude is completely consistent with the
proposed ultra--compact nature of 1H~1905+000. In order to fit observations the companion star has to be fainter
than an M2~V star (\citealt{2000asqu.book.....C}). Hence, a (hot) brown dwarf companion star such as that of
SAX~J1808.4--3658 (\citealt{2001ApJ...557..292B}) would also be consistent with the current constraints.

The low quiescent X--ray luminosity of this source shows that the difference in the quiescent luminosity of black
hole and neutron star SXTs found initially (e.g.~\citealt{2001ApJ...553L..47G}) may have been a selection effect
(see also \citealt{2005ApJ...635.1233T}). The observations reported here show that at least some of the neutron
star SXTs can be as faint as the faintest black hole SXTs in quiescence. The claim that the comparison between the
black hole and neutron star luminosity in SXTs in quiescence provides evidence for a black hole event horizon is
hence difficult to maintain. When comparing  Eddington scaled neutron star and black hole luminosities, as has been
done often in the literature, black hole SXTs are still less luminous than their neutron star counterparts. The
reasoning behind such a scaling stems from the notion that at orbital periods of the order of hours the
gravitational wave radiation driven mass transfer rate for neutron stars and black holes is roughly similar when
scaled to the Eddington rate {\it if the companion stars are main sequence stars} (\citealt{1999ApJ...520..276M}).
However, for ultra--compact systems (as 1H~1905+000 likely is) the companion star cannot be a main sequence star
making the Eddington scaling arbitrary. Furthermore, it is unclear whether the mass transfer rate set by the
orbital period and the instantaneous mass accretion rate have a one--to--one correspondence in quiescent systems. 

The low luminosity in quiescence can potentially be used together with the neutron star cooling theory to put
constraints on the presence of condensates in the neutron star core (\citealt{2004ARA&A..42..169Y}). Besides the
core cooling mechanism we can constrain the thermal relaxation time and hence thermal conductivity of the neutron
star crust. Since we do not detect the neutron star less than 20 years after the 11 year long outburst the neutron
star crust must have a thermal relaxation time less than 20 years. However, presently the data does not allow us to
distinguish between crustal conductivity set by electron--phonon conductivity (\citealt{1995AstL...21..702B}) or by
electron--ion scattering (\citealt{1980SvA....24..303Y}), since the inclusion of Cooper--pair neutrino emission in
the crust alters the description considerably with respect to the status presented in \citet{2001MNRAS.325.1157U}
(e.g.~see \citealt{1999A&A...343..650Y}). Currently, less than 20 years after the outburst, the quiescent
luminosity is determined by the core cooling processes again.  However, to estimate the core temperature the
average mass accretion rate over the last 10$^4$-10$^5$ year has to be known (\citealt{2001ApJ...548L.175C}). This
we can estimate from binary evolution theory. A primer for the evolutionary state of the low--mass X--ray binary
(outside Globular Clusters) is the orbital period. In case of 1H~1905+000 the orbital period is unknown. However,
as explained above, it is likely that 1H~1905+000 is an ultra--compact X--ray binary. The mass accretion rate for
an ultra--compact system depends on the exact orbital period and the nature and age of the companion star but it
can be as low as $\sim 10^{-13}$ M$_\odot$ yr$^{-1}$ if the system is $\sim$10 Gyr old and if the orbital period is
60--90 minutes (e.g.~\citealt{verbunt1995}, \citealt{2003ApJ...598.1217D}, see also \citealt{2006MNRAS.366L..31K}).
If indeed the time averaged mass accretion rate is this low, the current limit on the luminosity does not provide a
constraint on the core cooling. However, at such low mass transfer rates it might be difficult to feed an outburst
that lasts longer than 11 years where the X--ray luminosity is $\sim 4\times10^{36}$ erg s$^{-1}$ as in the case of
1H~1905+000.\footnote{1H~1905+000 is only the third transient known to date that returned to quiescence after a
long--duration outburst.} In order to sustain such an accretion rate for 11 years the neutron star should have
accreted $\approx4\times10^{-9}$ M$_\odot$. If the mass transfer rate is 10$^{-13}$/$10^{-12}$ M$_\odot$ yr$^{-1}$
then it would take at least 3.5$\times10^{4}$/3.5$\times10^3$ year to build--up the accretion disc. In both cases
this is a significant fraction of the core heating timescale. \citet{1998ApJ...504L..95B} showed that in such a
case the fraction of the heat released deep in the crust that is used to heat the core is smaller than 1. Assuming
that the fraction is not less than 0.1 in 1H~1905+000 then, following \citet{1998ApJ...504L..95B} we derive that
for a time averaged mass accretion rate larger than $2\times10^{-12}$ M$_\odot$ yr$^{-1}$, enhanced neutrino
emission  processes must be operating in the core.

Next, we discuss the possibility that systematic effects make 1H~1905+000 appear faint in quiescence whereas
the true luminosity in quiescence is significantly higher. We shall show, however, that this is unlikely.
If the distance is systematically underestimated the source luminosity may be higher. However, the distance
for 1H~1905+000 is derived from the observed radius expansion burst peak flux. \citet{2003A&A...399..663K}
have calibrated this distance estimation method comparing the distances derived from the burst peak fluxes
of sources in globular clusters to the accurately known distances of those globular clusters. From their
work it can be seen that if the composition of the burning material is known (hydrogen or helium rich
burning material) the peak burst flux gives a reliable estimate of the distance (to within approximately 15
per cent). In ultra--compact systems hydrogen is likely depleted (\citealt{1986ApJ...304..231N}; see
\citealt{2002ApJ...565.1107P} for a possible binary evolutionary scenario leading to ultra--compact systems
with some hydrogen still present). Therefore, it is save to assume that the radius expansion burst peak
flux corresponds to the helium Eddington limit luminosity. In case of 1H~1905+000 this helium radius
expansion burst limit on the distance is 10 kpc (see \citealt{2004MNRAS.354..355J}). If however, hydrogen
was present in the burning material the distance becomes smaller, making the upper limit on the quiescent
luminosity more stringent still. The interstellar extinction, N$_H$, towards 1H~1905+000 is low compared to
that found for most other low--mass X--ray binaries. From Einstein observations obtained when the source
was in outburst Christian \& Swank (1997) determined that the N$_H$ to 1H~1905+000 is
$(1.9\pm0.2)\times10^{21}$~cm$^{-2}$. Typically, for many SXTs it is found that the interstellar extinction
probed by N$_H$ is somewhat higher in outburst than in quiescence (\citealt{2004MNRAS.354..355J}). For that
reason it is unlikely that the neutron star quiescent luminosity is much higher but that the source is
hidden from our view due to an interstellar extinction that is much larger than measured in outburst.

Previous observations of several accretion--powered millisecond X--ray pulsars in quiescence have also
shown that many of those sources have a low quiescent luminosity (e.g.~SAX~J1808.4--3658,
\citealt{2002ApJ...575L..15C}, XTE~J0929--314 and XTE~J1751--305, \citealt{2005ApJ...619..492W},
XTE~J1807--294, \citealt{2005A&A...434L...9C}). However, except for SAX~J1808.4--3658, the distance
estimates for these systems are uncertain which makes the quiescent luminosity uncertain as well. Although,
in order for these systems to have a luminosity ${\rm \sim10^{33} erg s^{-1}}$, i.e.~similar to that
observed for Aql~X--1 and XTE~J1709--267 (\citealt{2002ApJ...577..346R}, \citealt{2003MNRAS.341..823J}),
the distance has to be unrealistically large, e.g.~$\sim40$ kpc for XTE~J0929--314. The planned deep (300
ksec) {\it Chandra} X--ray observation of 1H~1905+000 will provide new constraints on or a measurement of
the quiescent neutron star luminosity in this system.

\section*{Acknowledgments}  

\noindent Support for this work was provided by NASA through Chandra Postdoctoral Fellowship grant number
PF3--40027 awarded by the Chandra X--ray Center, which is operated by the Smithsonian Astrophysical Observatory for
NASA under contract NAS8--39073. PGJ further acknowledges support from NASA grant GO4-5033X fund number 16617404.
PGJ, CGB and GN acknowledge support from the Netherlands Organisation for Scientific Research. This research was
(partially) based on data from the ING Archive. We further acknowledge Tom Marsh for use of his software package
\textsc {molly}, Sergio Ilovaisky for providing the original pdf--image of the V--band finder chart of 1H~1905+000
published in Chevalier, Ilovaisky \& Charles (1985), and Frank Verbunt for comments on a earlier version of the
manuscript. The authors would like to thank the referee for his/her comments which improved the manuscript.

\end{document}